\documentclass[aps,prl,twocolumn,groupadress]{revtex4}
\usepackage{setspace}
\usepackage{graphicx}
\usepackage{amsmath}
\usepackage{amssymb}
\usepackage{epsfig}
\usepackage{dcolumn}
\usepackage{amsmath}
\usepackage{graphicx}

\begin{document}

\title{
Observation of Anomalous Phonon Softening in Bilayer Graphene
      }

\author{Jun Yan$^{1}$}
\email{jy2115@columbia.edu}
\author{Erik A. Henriksen$^{1}$}
\author{Philip Kim$^{1}$}
\author{Aron \surname{Pinczuk}$^{1, 2}$}

\affiliation{ ${}^1$Department of Physics, Columbia University, New
York, NY 10027, USA
\\ ${}^2$ Department of Applied Physics and Applied Mathematics, Columbia University, New York, NY 10027, USA
}

\begin{abstract}
The interaction of electron-hole pairs with lattice vibrations
exhibits a wealth of intriguing physical phenomena. The Kohn anomaly
is a renowned example where electron-phonon coupling leads to
non-analytic phonon dispersion at specific momentum nesting the
Fermi surface \cite{1}. Here we report evidence of another type of
phonon anomaly discovered by low temperature Raman spectroscopy in
bilayer graphene where the charge density is modulated by the
electric field effect. This anomaly, arising from charge-tunable
modulations of particle-hole pairs that are resonantly coupled to
lattice vibrations, is predicted to exhibit a logarithmic divergence
in the long-wavelength optical-phonon energy. In a non-uniform
bilayer of graphene, the logarithmic divergence is abated by charge
density inhomogeneity leaving as a vestige an anomalous phonon
softening. The observed softening marks the first confirmation of
the phonon anomaly as a key signature of the resonant
deformation-potential electron-phonon coupling. The high sensitivity
of the phonon softening to charge density non-uniformity creates
significant venues to explore the interplay between fundamental
interactions and disorder in the atomic layers.

\end{abstract}

\maketitle

\newcommand{\oph}{$\omega_{ph}$}
\newcommand{\EF}{$E_F$}
\newcommand{\Eeh}{$E_{e-h}$}
\newcommand{\hoG}{$\hbar\omega_G$}
\newcommand{\dn}{$\delta n$}
\newcommand{\incl}{\includegraphics}

Recently, the ability to use the electric field effect (EFE) to
continuously dope large densities of electrons or holes into
graphene \cite {2} has led to general interest in the deformation
potential coupling of optical phonons with charge carriers in the
extreme two-dimensional(2D) atomic limit \cite {3, 4, 5, 6, 7}, and
to experimental observations of charge-tunable phonon energy and
lifetime in single layer graphene \cite {8, 9, 10}. In prior
EFE-Raman experiments, however, a logarithmically-divergent
phonon-energy anomaly\cite {3} predicted by the dynamical
perturbation theory was not observed. The puzzling absence of the
phonon anomaly was attributed to the presence of large
non-uniformity of the charge density in monolayer graphene samples
\cite {8}. This interpretation is subsequently supported by the
observation of asymmetric phonon line-shape in the EFE-Raman spectra
at low doping levels\cite{11}.

\begin{figure*}
\incl[width=20.5cm, width=0.75\linewidth]{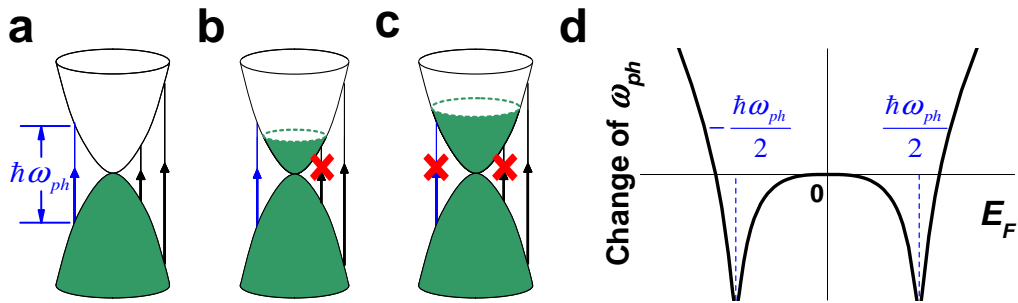}
\caption{\label{concept} {\bf Illustration of the phonon anomaly
concept.} {\bf a-c}, Vertical interband electron-hole pair
transitions in a gapless 2D semiconductor with 3 different Fermi
levels. Regions with green shading are filled with electrons. The
transition indicated by the blue arrow is the resonance with the
long-wavelength optical phonon. {\bf d}, Predicted change of phonon
energy as a function of the Fermi energy. The two phonon anomalies
show up at $E_F=\pm\hbar$\oph/2.}
\end{figure*}

In this paper we demonstrate that the phonon anomaly, while proposed
initially in single layer graphene \cite{3}, is a generic feature of
tunable resonant electron-phonon coupling when the particle-hole
pair energy is close to the phonon energy (Fig.1). We report
experimental observation of the anomalous phonon softening, a clear
signature of such phonon anomaly, in bilayer graphene. The
observation of the resonant phonon softening is facilitated in
bilayer graphene. The reason is that the coupling between the two
graphene layers results in a nearly parabolic dispersion and
relatively large density of states near the vanishing band gap,
making the phonon anomaly robust even in the presence of relatively
large charge density non-uniformity.

Rather than being a special property unique to the monolayer
graphene, the anomaly is anticipated to occur in any multi-band
electron system where the zero momentum optical phonon can create
resonant electron-hole pairs across the conduction and valence
bands. Figure 1 illustrates the concept of the phonon anomaly in a
2D gapless semiconductor with particle-hole symmetry as a model
system. A system with similar low energy band structure is in fact
bilayer graphene, in which the two valleys of conduction and valence
bands are centered at K and K' corner points of the Brillouin zone
\cite {12}. The deformation-potential interaction couples vertical
electron-hole interband excitations with long-wavelength optical
phonons in graphene \cite {13, 14}. This coupling contributes to the
renormalization of the phonon energy \cite {15}. Considering the
electron-phonon interaction within 2nd order time-dependent
perturbation theory \cite {15}, combined with the Pauli exclusion
principle, we obtain the change of phonon energy \oph with tuning of
the Fermi energy \EF:

\begin{widetext}
\begin{equation}
\centering \hbar\omega_{ph}(E_F)-\hbar\omega_{ph}(E_F=0) \sim
-\lambda\int_0^{2|E_F|}
dE_{e-h}\frac{2E_{e-h}}{\hbar\omega_{ph}{}^2-E_{e-h}{}^2} \sim
  \lambda ln|1-\frac{2|E_F|}{\hbar\omega_{ph}}|
\end{equation}
\end{widetext}
where \Eeh is the energy of an electron-hole pair, $\lambda$ is the
electron-phonon coupling parameter with dimension of energy. Because
of the resonant denominator
$\frac{1}{\hbar\omega_{ph}{}^2-E_{e-h}{}^2}$ in equation (1), the
perturbative contributions to \oph have a sign change at
\Eeh=$\hbar$\oph {}and are markedly enhanced near the resonance (the
interband transition indicated with a blue arrow in Fig. 1a). When
$|E_F|$ is increased, the positive and negative enhanced
perturbations to \oph{} are switched off due to restrictions placed
by the Pauli Principle (illustrations for the case of electron
doping are shown in Fig. 1b,c). For this reason, the modulation of
carrier density in the system results in marked changes in the
optical phonon energy around $|E_F| = \hbar$\oph/2, as shown in Fig.
1d.

Note that in the derivation of equation (1), no assumptions of
system dimensionality or band dispersion curvature were made. We
expect the phonon anomaly in principle to show up in 1D system like
metallic or near zero gap semiconducting carbon nanotubes \cite {16,
17}, graphene nanoribbons \cite {18}, 2D system like graphene thin
films, 3D system like silicon and germanium \cite {19, 20}. In fact,
like in single layer graphene, similar log-divergence was predicted,
albeit not observed, in the studies of chemically doped silicon
\cite {19}.

\begin{figure}
\centering \epsfig{figure=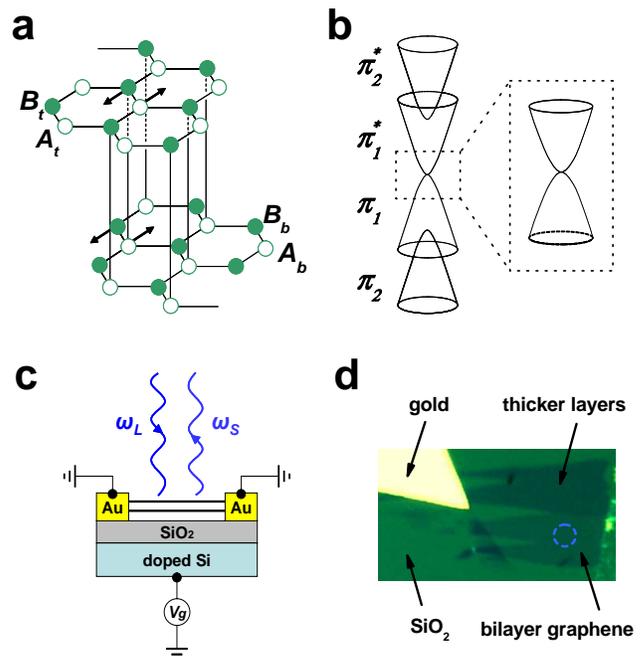, width=0.96\linewidth,clip=}
\caption{\label{bilayer} {\bf Bilayer graphene and the EFE-Raman
setup.} {\bf a}, Bilayer graphene lattice with 4 carbon atoms $A_t,
B_t, A_b, B_b$ in its unit cell. Black arrows indicate carbon atom
motion in the G band lattice vibration. {\bf b}, $\pi$ bands of
bilayer graphene near the K and K' corner points of the Brillouin
zone. The low energy region is zoomed out to illustrate that the
dispersion is drastically different from single layer graphene
\cite{12}. {\bf c}, Schematic drawing of the experimental setup. The
two black lines in-between the gold contacts represent bilayer
graphene. {\bf d}, Optical image of the bilayer graphene device. The
dashed blue circle is the position of the laser spot.} \label{fig2}
\end{figure}
\begin{figure*}
\incl[width=18.5cm, width=0.8\linewidth]{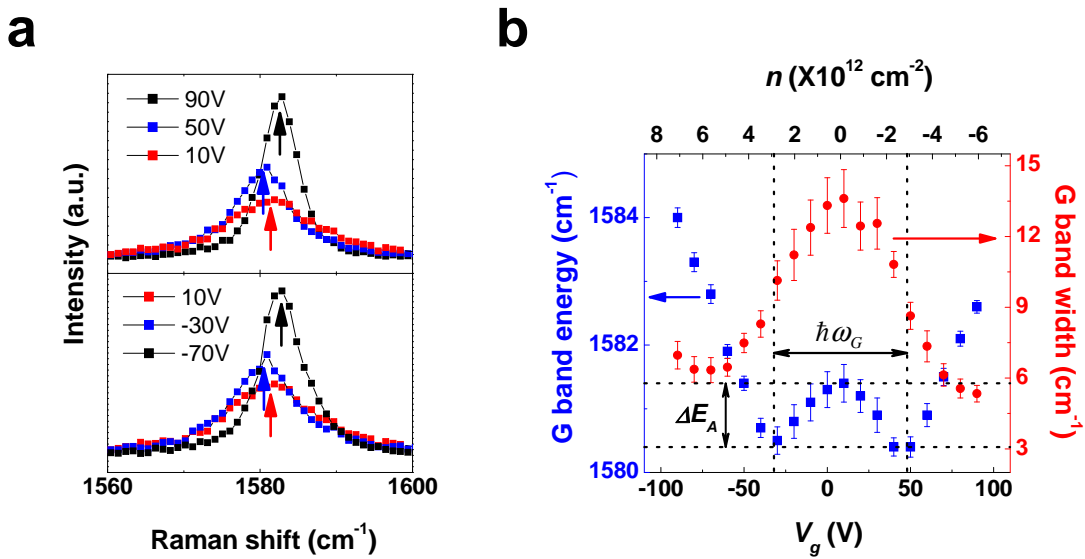}
\caption{\label{Raman} {\bf Evolution of the G band of bilayer
graphene with charge doping.} {\bf a}, G band at 5 different gate
voltages. The arrows indicate the phonon peak positions. Spectra
were taken at 12K in Helium gas environment. {\bf b}, Energy and
line-width of G band extracted from spectra in panel a. Two phonon
anomalies are clearly resolved in the phonon energy evolution (blue
squares).}
\end{figure*}

In the present work, we use EFE to tune \EF{} to access the resonant
electron-phonon interaction condition of the phonon anomaly in
bilayer graphene. Low temperature Raman spectroscopy was employed to
detect the evolution of the zero momentum optical phonon, known as G
band in graphite and graphene thin films. Its energy \hoG $\approx$
1580 cm$^{-1}$ $\approx$ 196 meV. Bilayer graphene is an intriguing
electronic system in which transport measurements have revealed
unconventional quantum Hall effect and Berry's phase of 2$\pi$ \cite
{21}. It has 4 carbon atoms in its unit cell. The locations of them
are $A_t $ $B_t$ in the top layer, $A_b$ $B_b$ in the bottom layer
(Fig. 2a). The $p_z$ orbital of these atoms forms 4 $\pi$ bands in
Fig. 2b.

Figure 2c schematically shows our experimental set-up. A back gate
voltage $V_g$ is applied across a thin layer of SiO$_2$ dielectric
sandwiched by bilayer graphene and doped silicon to induce charge
carriers in the sample. The gating efficiency is about $7.2 \times
10^{10}$ cm$^{-2}$/Volt \cite {22}. For Stokes Raman scattering,
$\omega_{Stokes} = \omega_L - \omega_S$, where $\omega_L$ and
$\omega_S$ are frequencies of the incident and scattered light,
respectively. The experiment is performed with the sample mounted
inside a variable temperature cryostat with optical access. Data
shown in this work are taken at 12 K.

The evolution of bilayer G band with the gate voltage $V_g$ is
displayed in Fig. 3. The EFE induced changes in the spectra are
nearly symmetric about $V_g$ = 10V (Fig. 3a). As in single layer
graphene \cite {8}, this symmetry, which determines the
charge-neutral point of the sample, reflects the underlying
particle-hole symmetry in the band structure of bilayer graphene.
Another important feature in the data is that, the phonon bands have
smaller line width at large charge doping with electrons or holes,
indicating longer lifetime. This change of width is due to Landau
damping of the G-phonon into electron-hole pairs when $|E_F|<$
\hoG/2 at small charge doping. Similar effects were observed in EFE
tuned Raman spectra in single layer graphene \cite {8}.

\begin{figure*}
\incl[width=18.5cm,angle=270, bb=29 27 578
768,width=0.8\linewidth]{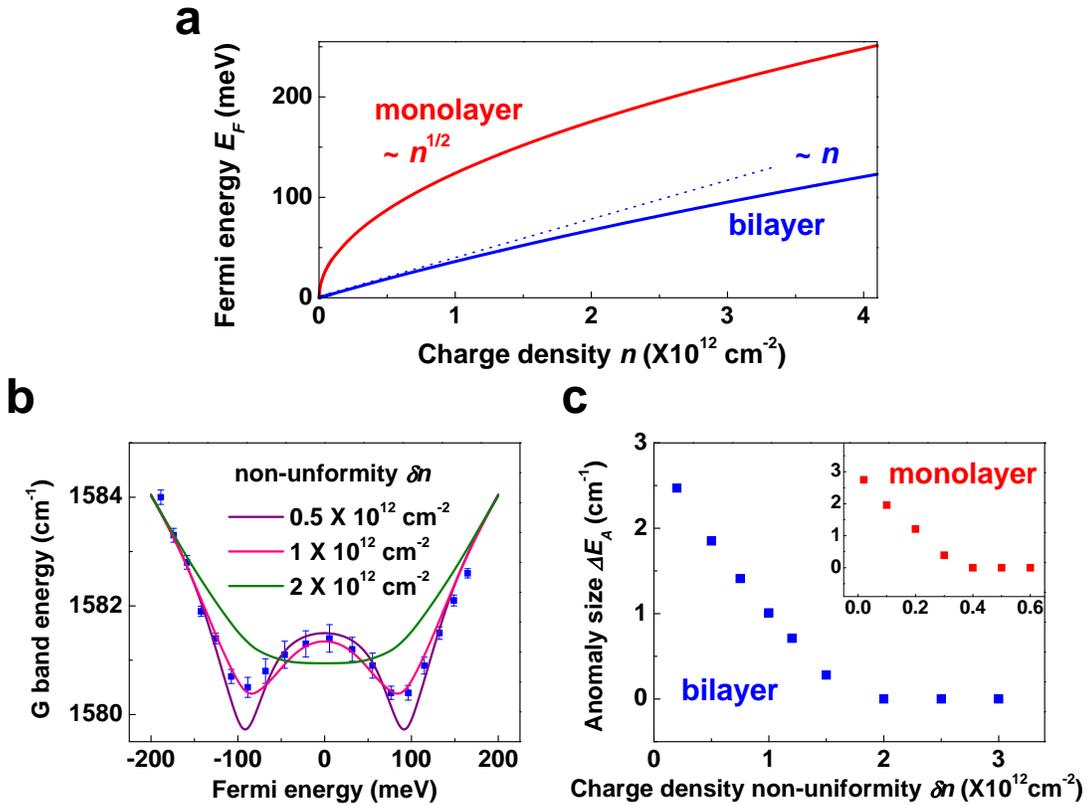} \caption{\label{anomaly} {\bf
Broadening of the phonon anomaly by charge density non-uniformity.}
{\bf a}, Comparison of $E_F\sim n$ relation in bilayer and monolayer
graphene. The Fermi energy goes up much faster with charge density
in the monolayer. {\bf b}, Fits of the evolution of the G phonon
energy with \EF. The best fit has charge density non-uniformity
$\delta n$ = 1$\times 10^{12}$ cm$^{-2}$. Blue squares are
experimental data taken from Fig. 3b. {\bf c}, Phonon anomaly
$\Delta E_A$ as a function of charge density non-uniformity in the
bilayer (main panel) and the monolayer (inset). The anomaly is more
robust in bilayer graphene.}
\end{figure*}

While the observed G phonon line width evolution in bilayer is
similar to that of a single layer, changes in the phonon energy are
drastically different. In single layer graphene, G phonon frequency
exhibits only one minimum as the EFE tuned charge density $n$ passes
through the charge-neutral Dirac point, and the phonon monotonically
stiffens with increasing $|n|$ \cite {8, 9, 10}. In contrast, when
charge carriers are added into the bilayer sample, \hoG{} first
decreases for smaller $|n|$ and then increases at larger doping
(square symbols in Fig. 3b). As expected from the particle-hole
symmetry in the system, two distinct minima are clearly resolved in
\hoG($n$). These minima indicate that G phonon stiffness changes
from softening to hardening as carrier density increases.

Non-monotonic changes of phonon energy as a function of carrier
density is quite unusual as it has never been observed before. In
chemically doped silicon \cite {19, 20} and EFE doped monolayer
graphene \cite{8, 9, 10}, the phonon stiffness always changes
monotonically with the carrier density. In this context, the double
minima shown in Fig. 3b is quite anomalous. In the following, we
ascribe that the observed minima are indeed manifestations of the
general phonon anomaly described in equation (1) above. First, we
note the similarities between Fig. 1d and Fig. 3b, suggesting that
the phonon energy minima correspond to \EF= $\pm$ \hoG/2 in bilayer
graphene. Furthermore, between the two minima, the phonon line width
(red dots in Fig. 3b) is larger than outside of the minima,
indicating that Landau damping of the phonon into resonant
electron-hole pair transition (blue arrows in Fig. 1a,b) is allowed
within this region \cite {8}. This is consistent with the fact that
the anomaly positions correspond to the hole (left minimum) and the
electron (right minimum) in the resonant electron-hole pair
respectively.

Since the phonon anomaly can only occur at a special Fermi energy,
charge density difference between these two minimal points $\Delta
n_A \approx 6\times10^{12}$cm$^{-2}$ is linked to the electronic
band parameters that determine the low energy dispersion of the
bilayer graphene. Within the tight-binding model \cite {23}, we
found from $\Delta n_A$ the inter-layer $A_b-B_t$ hopping energy
$\gamma_1 = 0.35\pm 0.06$eV. This value is in reasonable agreement
with $\gamma_1 =0.43\pm0.03$eV obtained from photoemission spectra
in epitaxial bilayer graphene \cite {24}.

We expect the phonon anomaly to show up in single layer graphene if
the sample quality is improved. In previous measurements
\cite{8,11}, single layer graphene samples contain large charge
inhomogeneity, which yields electron and hole puddles of size
$\delta n = 3-10\times 10^{11}$cm$^{-2}$. This inhomogeneity
corresponds to the mesoscopic Fermi energy broadening $\delta E_F
\sim$ 100 meV near the Dirac point in single layer graphene, a value
large enough to wash out the anomalous phonon softening completely.
In bilayer graphene, however, similar $\delta n$ results in  much
smaller $\delta E_F$, because bilayer $E_F$ changes much slower with
$n$ than the monolayer ($n$ versus $\sqrt{n}$ dependence in the low
density regime), as shown in Fig. 4a. For this reason, the phonon
anomaly is more robust and easier to observe in bilayer graphene.

To be more quantitative, we evaluated numerically the effect of
charge inhomogeneity on \hoG(\EF), assuming a Gaussian distribution
of the charge density $f(n)=\frac{1}{\sqrt{2\pi}\delta n}
e^{-(\frac{n-n_0}{\sqrt{2}\delta n})^2}$, where $n_0$ is the average
charge density in the sample, \dn{} represents the size of the
charge density non-uniformity. Figure 4b exemplifies several such
evolutions for different \dn. As expected, the larger $\delta n$ is,
the smaller is the phonon anomaly size $\Delta E_A$ = \hoG(\EF=0) -
\hoG(\EF=\hoG/2). Figure 4c displays $\Delta E_A$ as a function of
\dn. Comparing with the experimental observation, we estimate that
\dn{} $\approx1\times10^{12}$ cm$^{-2}$ for 1 cm$^{-1}$ $\Delta E_A$
in a bilayer. This size of charge density non-uniformity agrees well
with results obtained from other experimental methods \cite {25, 26,
27}. Note that similar size of charge density non-uniformity in
single layer graphene is enough to smooth out the phonon anomaly
completely (inset of Fig. 4c).

In conclusion, we have observed the anomalous softening of the
long-wavelength optical phonon in bilayer graphene. This striking
effect is a manifestation of the predicted logarithmic divergence
due to tunable resonant coupling of phonons with particle-hole
pairs. The broadening of the phonon anomaly is attributed to large
charge density non-uniformity in graphene, showing that EFE-Raman
spectroscopy can access fundamental interaction effects near the
charge neutral point of graphene layers even in the presence of an
inhomogeneous charge density distribution.

{\small We thank I. Aleiner, D. Basko and A. Millis for helpful
discussions. We acknowledge financial support from NSF
(CHE-0117752), the NYSTAR, and ONR (N000140610138). P.K.
acknowledges support from the FENA MARCO Center. A. P. is supported
by NSF (DMR-0352738) and DOE (DE-AIO2-04ER46133).

\textit{Note added. - } After completion of this work, we noticed a
theoretical paper \cite{28} on electron-phonon coupling in bilayer
graphene, where it was confirmed that the long-wavelength
optical-phonon energy again exhibits a logarithmic divergence as the
charge density in the bilayer is continuously tuned.

}

\end{document}